\documentclass[letter]{aa}
\usepackage{graphicx}
\usepackage{txfonts}
\usepackage{booktabs}
\usepackage{natbib}
\usepackage[utf8]{inputenc}
\usepackage{multirow}
\bibpunct{(}{)}{;}{a}{}{,} % to follow the A&A style

\newcommand{\swift}{\textit{Swift}}

\newcommand{\fermi}{\textit{Fermi}}

\usepackage[usenames,dvipsnames]{color}
\usepackage{colortbl,xcolor}

\begin{document}

% ------------------------------------------------------------------------
\title{The late-time afterglow of the extremely energetic short burst \\ 
GRB 090510 revisited}

\author{
A.~Nicuesa Guelbenzu\inst{1}, 
S.~Klose\inst{1},
T.~Kr\"uhler\inst{2,3,4},
J.~Greiner\inst{2},
A.~Rossi\inst{1},
D.~A.~Kann\inst{1},
F.~Olivares~E.\inst{2},
A. Rau\inst{2},
P.~M.~J.~Afonso\inst{2,5},
J.~Elliott\inst{2},
R.~Filgas\inst{2},
A.~K\"upc\"u~Yolda\c{s}\inst{6},
S.~McBreen\inst{7},
M. Nardini\inst{2},
P.~Schady\inst{2},
S. Schmidl\inst{1},
V.~Sudilovsky\inst{2},
A.~C.~Updike\inst{8,9,10},
\and
A.~Yolda\c{s}\inst{6}
}

\institute{Th\"uringer Landessternwarte Tautenburg, Sternwarte 5, 07778 Tautenburg, Germany\\
\email{ana@tls-tautenburg.de}
\and
Max-Planck-Institut f\"ur Extraterrestrische Physik, Giessenbachstra\ss{}e, 85748 Garching, Germany
\and
Universe Cluster, Technische Universit\"{a}t M\"{u}nchen, Boltzmannstra\ss{}e 2, 85748, Garching, Germany
\and
Dark Cosmology Centre, Niels Bohr Institute, University of
Copenhagen, Juliane Maries Vej 30, 2100 Copenhagen, Denmark
\and
American River College, Department of Physics and Astronomy, 4700 College Oak Drive,
Sacramento, CA 95841, USA
\and
Institute of Astronomy, University of Cambridge, Madingley Road CB3 0HA, Cambridge, UK
\and
School of Physics, University College Dublin, Dublin 4, Republic of Ireland
\and
Clemson University, Department of Physics and Astronomy, Clemson, SC 29634-0978, USA
\and 
CRESST and the Observational Cosmology Laboratory, NASA/GSFC, Greenbelt, MD 20771, USA 
\and
Department of Astronomy, University of Maryland, College Park, MD 20742, USA}

% ------------------------------------------------------------------------
\date{Received 2011 November 7; accepted XXXX}

\authorrunning{Nicuesa Guelbenzu et al.}

\titlerunning{GRB 090510}

% ------------------------------------------------------------------------
\abstract
% context 
{The \swift \ discovery of  the short burst GRB 090510 has raised considerable
attention mainly because of two reasons: first, it had a bright optical
afterglow, and second it is among the most energetic events detected so far
within the entire GRB population (long plus short).  The afterglow of GRB
090510 was observed with \swift/UVOT and \swift/XRT and evidence of a jet
break around 1.5~ks after the burst has been reported in the literature,
implying that after this break the optical and X-ray light curve should fade
with the same decay slope.}
% aims 
{As noted by several authors, the post-break decay slope seen in the UVOT data
is much shallower than the steep decay in the X-ray band, pointing  to a
(theoretically  hard to understand) excess of optical flux at late times. We
assess here the validity of this peculiar behavior.}
% methods
{We reduced and analyzed new afterglow light-curve data obtained with the
multichannel imager GROND. These additional $g'r'i'z'$ data were then
combined with the UVOT and XRT data to study the behavior of the afterglow
at late times more stringently.}
% results
{Based on the densely sampled data set obtained with GROND, we find that the
optical afterglow of GRB 090510 did indeed enter a steep decay phase starting
around 22~ks after the burst.  During this time the GROND optical light curve
is achromatic, and its slope is identical to the slope of the X-ray data. In
combination with the UVOT data this implies that a second break must have
occurred in the optical light curve around 22~ks post burst, which, however,
has no obvious counterpart in the X-ray band, contradicting the interpretation
that this could be  another jet break.}
% conclusions 
{The GROND data provide the missing piece of evidence that the optical
afterglow of GRB 090510 did follow a post-jet break evolution at late
times. The break seen in the optical light curve around 22~ks in combination
with its missing counterpart in the X-ray band  could be due to the passage of
the injection frequency across the optical bands, as already theoretically
proposed in the literature. This is  possibly the first time that this passage
has been  clearly seen in an optical afterglow. In addition, our results imply
that there is no more evidence for an  excess of flux in the optical bands at
late times.}     
{}


\maketitle

% ------------------------------------------------------------------------
\section{Introduction}

After the first GRB was discovered in 1967 (\citealt{Klebesadel1973}), GRB
research has evolved rapidly. In the early 1990s it became clear that GRBs
come in two flavors, long and short, with the borderline around 2~s
(\citealt{Kouveliotou1993}). Thanks to three generations of  high-energy
satellites, BeppoSAX (\citealt{Piro1998}), HETE-2 (\citealt{Ricker2002}), and
\swift\ (\citealt{Gehrels2004}), it is now known that long GRBs are linked to
the core collapse of massive stars (\citealt{WB2006}), while short bursts are
most likely linked to compact stellar mergers in all morphological types of
galaxies (\citealt{Nakar2007,Fong2010a}). Short bursts are much less frequently
observed than long GRBs so that our knowledge about short burst
progenitors is much less complete. 

Since mid-2007 our group operates the seven-band imager GROND mounted at the 2.2m
ESO/MPG telescope on La Silla, especially designed for GRB follow-up
observations (\citealt{Greiner2008}). Every observable burst is followed with
delay times down to 2.5 minutes between the GRB trigger and the first
exposure. 

\begin{figure*}[t]
\includegraphics[width=18.4cm,angle=0]{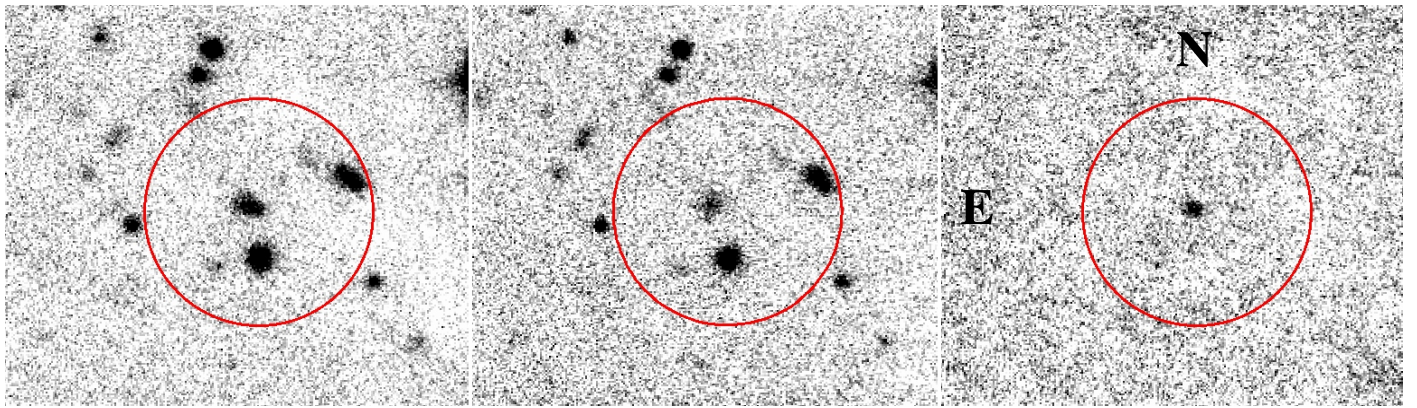}
\caption{GROND ($g^\prime r^\prime i^\prime z^\prime $) white band finding
chart of the afterglow of GRB 090510. $Left$: the afterglow plus its host
galaxy on the night of the burst between 22~ks and 36~ks after the
trigger. The afterglow flux dominates the western part of its host galaxy.
(North is up and east is to the left.) $Middle$: In the second night the
afterglow faded away, with only the host galaxy visible. $Right$: Image
subtraction between the first and the second epoch  white band image clearly
revealing the afterglow. The circle with a radius of $10^{\prime\prime}$,
centered on the position of the afterglow, is just drawn to guide the eye.}
\label{fig:090510fc}
\end{figure*}

GRB 090510 triggered \swift/BAT (\citealt{Hoversten2009}) and \fermi/GBM
(\citealt{Guiriec2009}) on 10 May 2009 at 00:23:00 UT, as well as \fermi/LAT at
00:23:01 UT (\citealt{Ohno2009}).  In the \swift/BAT energy window it had a
duration of $T_{90}\;\textrm{[15, 350~keV]}=0.3\pm0.1$~s
(\citealt{Hoversten2009GCNR218}).  \swift/XRT started observing the field
about 94~s after the trigger and the X-ray afterglow was immediately found
(\citealt{Hoversten2009}). \swift/UVOT began  observations shortly
after the XRT, and an optical afterglow candidate was also seen
(\citealt{Marshall2009GCN9332,Kuin2009GCN9342}), which was soon confirmed  by
the Nordic Optical Telescope (\citealt{Olofsson2009GCN9338}) and by GROND
(\citealt{Olivares2009GCN9352}). The redshift of its underlying host galaxy
was finally measured using VLT/FORS2 about 2.3 days after the trigger
($z$=0.903; \citealt{Rau2009GCN9353,McBreen2010}).

GRB 090510 is not only one of the few short bursts with a clear afterglow
detection in the optical bands, but it is also especially unique because it is
among the most energetic events detected so far in the entire GRB population
(long plus short).  In particular, a 31~GeV photon from this burst
(\citealt{Abdo2009}) is the second highest energy photon ever received from a
GRB (see figure 5 in \citealt{Piron2011}). Naturally, the  afterglow of GRB
090510 was of special interest, too. Remarkably, all studies of its afterglow
(see Sect.~\ref{discussion}) agree on one point: when compared to its X-ray
light curve, its computed late-time decay slope in the UVOT white band is
difficult to understand within the framework of the standard afterglow model.

Here we present additional photometry of the optical afterglow of GRB 090510
obtained with GROND from about 22~ks to 36~ks after the burst, leading to a
re-evaluation of its late-time evolution.

% ------------------------------------------------------------------------
\section{Observations and Data Reduction}

GROND started observing the field 6.2 hours after the burst and continued  for
3.5 hours.  Owing to visibility constraints from ESO/La Silla, GROND could not
be on target earlier. The following night, the field was observed again with
GROND for 1.5 hours. Data was reduced in a standard fashion via standard PSF
photometry using DAOPHOT and ALLSTAR tasks under IRAF (\citealt{Tody1993}),
similar to the procedure described in \citet{Thomas2008} and
\citet{Aybuke2008}.  Calibrations were performed against the
SDSS\footnote{http://www.sdss.org/dr7/}. Magnitudes were corrected for
Galactic extinction, assuming $E(B-V)=0.02$ mag \citep{Schlegel1998} and a
ratio of total-to-selective extinction of $R_V=3.1$.

%-------------------------------------------------------------------
\section{Results and discussion}

\subsection{The afterglow light curve \label{lightcurve}} 

The first night, the GRB host galaxy is clearly visible in the optical images,
with the afterglow light dominating its western part
(Fig.~\ref{fig:090510fc}). While on the first night the afterglow was detected
in $g^\prime r^\prime i^\prime z^\prime $ but not in $JHK_s$, and the second
night there was no sign of an afterglow in any band, except the host
galaxy. Image subtraction clearly reveals the afterglow between the first- and
the second-epoch of the combined $g^\prime r^\prime i^\prime z^\prime $ images
using
HOTPANTS.\footnote{http://www.astro.washington.edu/users/becker/hotpants.html}
Its coordinates measured against the USNO-B1 catalog are RA (J2000) =
22:14:12.53, Dec. = $-$26:34:59.0, with an error of 0\farcs2 in each
coordinate. The afterglow lies about 1\farcs2 west of the center of its host
galaxy (see also \citealt{McBreen2010}).

\begin{figure}[t]
\includegraphics[width=9.2cm,angle=0]{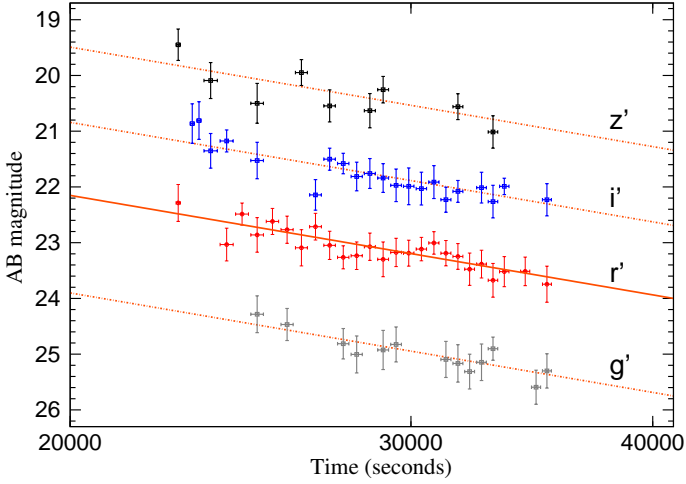}
\caption{GROND $g^\prime r^\prime i^\prime z^\prime $-band light curves of
the afterglow of GRB 090510 (from bottom to top). For reasons of clarity the
$g^\prime, i^\prime, z^\prime $ bands  were shifted by $+1.4, -1.0, -2.0$ mag,
respectively. The solid straight line is the best fit of the $r^\prime
$-band data, and the broken lines are its corresponding shifts to the other
bands. Shown are all data points with a 1$\sigma$ error of $\lesssim$0.35
mag (see Table~\ref{tab:log}).}
\label{fig:090510griz}
\end{figure}

During the first night, GROND detected the fading afterglow in all optical
bands (Fig.~\ref{fig:090510griz}, Table~\ref{tab:log}).  For this timespan,
from 22~ks to 36~ks, the $r^\prime $-band light curve can be fit by a single
power law with a slope of $\alpha_{\rm opt} = 2.37\pm0.29$ ($\chi^2_{\rm red}
=0.49$; 23 degrees of freedom).\footnote{In the following we use the standard
notation for the flux density, $F_\nu(t) \propto t^{-\alpha}\,
\nu^{-\beta}$.} This slope also fits the $g^\prime i^\prime z^\prime $ band
data; i.e., the evolution of the optical afterglow was achromatic.\footnote{
A joint fit leads to the same conclusion but has a slightly higher
$\chi^2_{\rm red}$ of 0.59.} Within its $1\sigma$ error, it also matches the
late-time decay  slope of the X-ray afterglow ($\alpha_{\rm X} = 2.18\pm
0.10$; \citealt{DePasquale2010}).  The obtained decay slope is substantially
different from what is reported by \cite{DePasquale2010} based on \swift/UVOT
data. 

% ----------------------------------------------------------------------
\subsection{The SED of the afterglow} 

The X-ray data for $t<20$ ks lead to a time-averaged spectral slope of
$\beta_{\rm X}=0.8\pm0.1$, while the X-ray data for $t>20$ ks give $\beta_{\rm
  X}=1.4\pm0.7$ (see the \swift/XRT Repository, \citealt{Evans2007a}). The XRT
data are  therefore consistent with having a spectral index of $\beta_{\rm
  X}=0.8$ throughout  the observations. This suggests that the cooling
frequency  $\nu_c$ lies above the X-ray band in the whole X-ray data set
(\citealt{DePasquale2010}). 

\begin{figure}[t]
\includegraphics[width=8.8cm,angle=0]{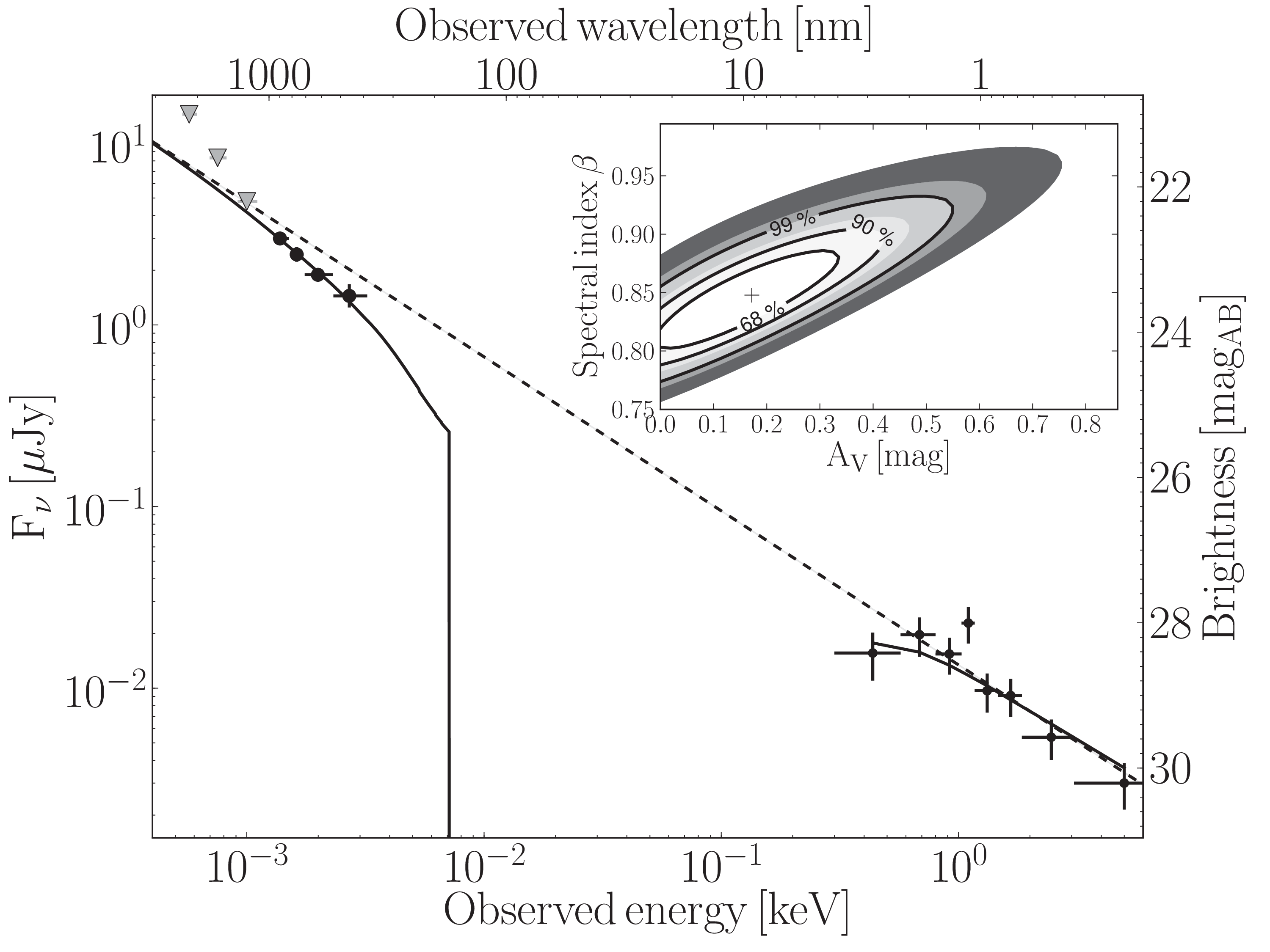}
\caption{\swift/XRT (\citealt{Evans2007a}) to optical/NIR (GROND) spectral
energy distribution of the afterglow of GRB 090510 at $t= 31$~ks after the
burst. The inset shows the $\beta_{\rm opt}$ vs $A_V^{\rm host}$ plane,
constraining their corresponding error bars. Filled triangles refer to the
GROND-observed NIR upper limits ($J$=22.2, $H$=21.6, and $K_s$=21.0),
filled circles to the observed optical magnitudes ($g',r',i',z'$).}
\label{fig:SED090510}
\end{figure}

Figure~\ref{fig:SED090510} shows the best SED fit from the optical to the
X-rays using $N_{\rm H}^{\rm Gal} = 1.7 \,\times\, 10^{20}$ cm$^{-2}$
(\citealt{Kalberla2005}).  For SMC dust and a redshift of $z=0.903$, it finds a
host galaxy extinction of $A_V^{\rm host} = 0.17_{-0.17}^{+0.21}$ mag, a gas
column density  of $N_{\rm H}^{\rm host} = 0.05_{-0.05}^{+0.15}
\,\times\,10^{22}$ cm$^{-2}$, and a spectral slope of $\beta_{\rm
  opt}=0.85\pm0.05$ ($\chi^2$/d.o.f.=0.93). 

% ------------------------------------------------------------------------
\subsection{What the second light curve break represents \label{discussion}}

As pointed out by several authors, the optical light-curve fit based on UVOT
data is difficult to understand when compared to the X-ray band.  A suggested
post-break decay slope of $\alpha_{\rm opt}\sim1.1$ (\citealt{DePasquale2010})
is very shallow when compared  to the  corresponding X-ray light-curve decay
($\alpha_{\rm X}=2.18\pm0.10$), implying that the optical bands show an excess
of flux at late times (\citealt{Corsi2010,DePasquale2010,He2011,Kumar2010,
Panaitescu2011}). 

At first we note that the steep decay of the optical flux seen by GROND
($\alpha_{\rm opt} = 2.37\pm0.29$; Sect.~\ref{lightcurve}) cannot be explained
as pre-jet break evolution. In our data base of afterglow light curves with a
well-observed pre- and post-jet break evolution (\citealt{Kann2010,Kann2011}),
we do not have a single case where the pre-break decay slope is 
as steep as that. We
conclude that at the time when the optical afterglow was monitored by GROND
the jet-break had already occurred, and the evolution of the afterglow was in
the post-jet break decay phase, confirming the finding of
\cite{DePasquale2010} based on the X-ray light curve. Second, in the GROND
$g^\prime r^\prime i^\prime z^\prime $ light curve data there is no evidence of
any break, and the decay is achromatic. The UVOT data then show
(\citealt{DePasquale2010}, their figure 1) that a break must have occurred
shortly before GROND started observing.

\begin{figure}[t]
\includegraphics[width=9.2cm,angle=0]{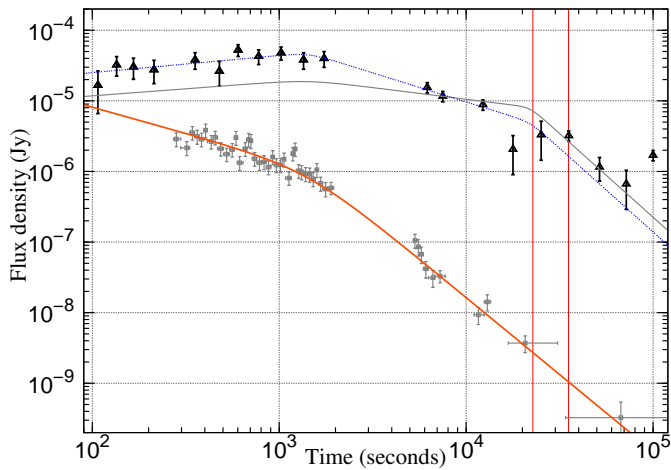}
\caption{Best fit of the UVOT \emph{white}-band data (black triangles, taken
from \citealt{DePasquale2010}, their figure 1) for two models 
that require a second break in the UVOT light curve at 22~ks (see
Sect.~\ref{discussion}). Also shown are the X-ray data, shifted in flux
density by a factor of 10$^4$. The two vertical lines highlight the time
interval in which GROND was observing.} 
\label{fig:090510.UVOT}
\end{figure}

Using the data published in \cite{DePasquale2010}, we fitted the UVOT
white-band  magnitudes again. At first we assumed a double-broken power law
(for the procedure see \citealt{Schulze2011}) with  fixed break times at
$t_{\rm b1}=1.4$ ks (based on the X-ray data, \citealt{DePasquale2010}) and
$t_{\rm b2}=22$ ks. Using a smoothing parameter of $n_1=n_2=10.0$ (see
\citealt{Beuermann1999}) for the first and second breaks, respectively,  and a
late-time decay slope of $\alpha_3=2.4$ (Sect.~\ref{lightcurve}), this gives
an early-time slope of $\alpha_1=-0.2\pm0.2$ and $\alpha_2=0.8\pm0.1$
(Fig.~\ref{fig:090510.UVOT}, blue  dashed  line). A relatively sharp break at
$t_{\rm b2}$ is required (defined by $n_2$) since the GROND data do not show
evidence  of any curvature in the light curve (Fig.~\ref{fig:090510griz}).
The UVOT two data points at 18 ks and 100 ks are strong outliers,
however.\footnote{In the second night GROND was observing between 116 ks and
  122 ks after the burst. We do not see evidence of rebrightening.} This
solution suggests we interpret $\alpha_2$ as a normal pre-jet break decay
slope. There is, however, no clear evidence of a corresponding (i.e.,
achromatic) break in the X-ray light curve, contradicting this interpretation
and in this way not affecting the generally excepted idea of a jet break time
already around 1.4 ks after the burst
(\citealt{Corsi2010,DePasquale2010,He2011,Kumar2010,Panaitescu2011}).

Another approach for fitting the UVOT data is suggested by a model discussed by
\cite{DePasquale2010} and \cite{Kumar2010}. When interpreting the XRT/UVOT
data, these authors point out that the flat UVOT light curve decay for
$t\gtrsim$ 1 ks ($\alpha_{\rm opt}\sim1.1$) can be understood if, at the time
when UVOT was observing, the injection frequency was (still) above the optical
bands ($\nu_{\rm opt} < \nu_m$) and the afterglow was in the post-jet-break
phase. While theoretically this suggests a decay slope of $\alpha$=1/3 (e.g.,
\citealt{Zhang2004IJMPA}),  these authors argue that possibly  the crossing of
$\nu_m$ through the  UVOT bands affected the measured decay slope, making it
flatter. In addition, these authors note that the UVOT light curve does not
show evidence of any steepening to a decay slope with $\alpha = p$,  where $p$
is the power-law index of the electron distribution function, a steepening
that is expected once $\nu_m$ has passed through the optical bands. The GROND
data now suggest re-evaluating this idea, since the expected steepening to
$\alpha = p$ is indeed seen in the data but was  originally not clearly
evident in the sparse UVOT data set. 

When following this model, a possible fit of the UVOT data with a
double-broken power law is also shown in Fig.~\ref{fig:090510.UVOT} (gray
line). It uses fixed $\alpha_1=-0.2, \alpha_2=1/3$, and $\alpha_3=2.4$, fixed
break times as mentioned before, as well as $n_1=n_2=10$. While this fit
underpredicts the UVOT optical flux for $t<$2~ks by a factor of $\sim$2, for
$t>2$ ks it reasonable agrees with the observational data. In
particular, there is no more evidence of any excess of flux in the UVOT bands
at late times (except for the UVOT data point at 100~ks; see  footnote
\#5). Clearly, the underprediction of the optical flux at very early times is
a shortcoming of this approach. It remains open whether the very early optical
flux could have been affected by rebrightening episodes, similar to what has
been seen in, e.g., the afterglow of GRB 080928 (\citealt{Rossi2011}).

Is the second light curve break definitely caused by the passage of $\nu_m$
across the optical bands? At least one shortcoming of this interpretation
could be that, if $\nu_m\sim$3~eV at $t$=20~ks, then this predicts a
relatively high value for the injection frequency at very early times, notably
higher than suggested by detailed numerical models of the afterglow
(\citealt{Kumar2010}). This issue cannot be solved here. On the other hand, if
the second light-curve break were the classical jet break, this would call for
a complete re-evaluation of the afterglow parameters, and we would be
confronted with the problem of a very sparse X-ray data set around $t$=20~ks,
which would make such an approach even more speculative.

% ------------------------------------------------------------------------
\section{Summary and conclusions}

We have presented GROND multichannel data of the optical afterglow of GRB
090510 obtained between 22~ks and 36~ks after the burst.  These data suggest
that, while GROND was observing, the afterglow was in the post-jet break decay
phase with a slope of $\alpha\sim2.4$. In combination with \swift/UVOT data,
this implies that, in addition to a break at 1.4 ks, a  second break  occurred
in the optical light curve around 22~ks after the burst. The lack of any evidence of a
corresponding break in the X-ray light curve at 22~ks disfavors the idea that
this is a jet break.   Following the discussion in \cite{DePasquale2010} and
\cite{Kumar2010}, this second break  could be understood however as 
 the passage of the injection frequency $\nu_m$ across the optical bands,
when the afterglow was in the post-jet break decay phase. Furthermore, we find
that the GROND data resolve the original issue of a potential excess of flux
in the optical bands at late times. The late-time decay slope in the optical
bands after 22~ks (i.e., after the passage of $\nu_m$) is, within the errors,
identical to the slope of the X-ray light curve, as expected for a post-jet
break evolution. We conclude that there is no longer any evidence of an excess
of flux in the optical bands at late times. After 22~ks, the evolution of the
afterglow was achromatic from the optical to the X-ray band.

% -------------------------------------------------------------------------
\begin{acknowledgements}
A.N.G., D.A.K., and S.K. acknowledge support by grant DFG Kl 766/16-1. A.N.G.,
A. R., D. A. K., and A.U. are grateful for travel funding support through the MPE.
T.K. acknowledges funding  by the DFG cluster of excellence 'Origin and
Structure of the Universe', F.O.E. funding of his Ph.D. through the DAAD,
M.N. support by DFG grant SA 2001/2-1 and P.S. by DFG grant SA 2001/1-1. Part
of the funding for GROND (both hardware and personnel) was generously granted
by the Leibniz-Prize to G. Hasinger (DFG grant HA 1850/28-1). This work made
use of data supplied by the UK Swift science data center at the University of
Leicester. 
We thank the referee for a rapid reply and a careful reading of the
manuscript.
\end{acknowledgements}

%%%%%%%%%%%%%%%%%%%%%%%%%%%%%%%%%%%%%%%%%%%%%%%%%%%%%%%%%%%%%%%%%%%%%%%%%%%
%                               Bibliography
%%%%%%%%%%%%%%%%%%%%%%%%%%%%%%%%%%%%%%%%%%%%%%%%%%%%%%%%%%%%%%%%%%%%%%%%%%%

\bibliographystyle{aa}
\bibliography{Nicuesa.2011.bib}

\Online

\begin{appendix}

\section{Afterglow photometry} 

\begin{table}[!htb]
\caption[]{Log of the GROND observations, given in the AB system.
Data are not corrected for Galactic extinction.}
\renewcommand{\tabcolsep}{5.0pt}
\begin{center}
\begin{tabular}{lllll}
\toprule
  Time (s)   &  $g'$             & $r'$              &  $i'$             &  $z'$            \\
\midrule
       22299 &   --              &  22.01$\pm$  0.38 &   --              &  21.73$\pm$  0.56 \\
       22401 &   --              &   --              &   --              &  21.27$\pm$  0.40 \\
       22503 &   --              &  22.09$\pm$  0.38 &   --              &   --              \\
       22609 &   --              &   --              &  21.85$\pm$  0.40 &  21.41$\pm$  0.40 \\ 
       22743 &   --              &  22.29$\pm$  0.33 &   --              &  21.45$\pm$  0.28 \\
       22931 &   --              &  22.89$\pm$  0.57 &  22.16$\pm$  0.38 &   --              \\
       23127 &  22.88$\pm$  0.56 &  22.73$\pm$  0.51 &  21.86$\pm$  0.35 &   --              \\
       23313 &   --              &  22.91$\pm$  0.53 &  21.81$\pm$  0.34 &   --              \\
       23639 &  22.90$\pm$  0.36 &   --              &  22.35$\pm$  0.31 &  22.09$\pm$  0.32 \\
       24093 &   --              &  23.03$\pm$  0.29 &  22.18$\pm$  0.20 &  22.40$\pm$  0.45 \\
       24540 &   --              &  22.49$\pm$  0.20 &  22.79$\pm$  0.45 &   --              \\
       24984 &  22.88$\pm$  0.33 &  22.86$\pm$  0.31 &  22.53$\pm$  0.33 &  22.50$\pm$  0.36 \\
       25443 &   --              &  22.62$\pm$  0.23 &  22.86$\pm$  0.39 &   --              \\
       25889 &  23.07$\pm$  0.29 &  22.77$\pm$  0.24 &  22.98$\pm$  0.41 &  22.95$\pm$  0.45 \\
       26335 &  23.45$\pm$  0.43 &  23.09$\pm$  0.32 &   --              &  21.95$\pm$  0.23 \\
       26780 &  23.86$\pm$  0.50 &  22.71$\pm$  0.24 &  23.14$\pm$  0.27 &   --              \\
       27234 &  23.54$\pm$  0.37 &  23.05$\pm$  0.25 &  22.50$\pm$  0.20 &  22.55$\pm$  0.29 \\
       27679 &  23.41$\pm$  0.27 &  23.26$\pm$  0.21 &  22.58$\pm$  0.18 &   --              \\
       28125 &  23.61$\pm$  0.33 &  23.23$\pm$  0.25 &  22.81$\pm$  0.26 &  22.88$\pm$  0.37 \\
       28569 &   --              &  23.07$\pm$  0.24 &  22.76$\pm$  0.27 &  22.63$\pm$  0.31 \\
       29024 &  23.52$\pm$  0.35 &  23.30$\pm$  0.31 &  22.84$\pm$  0.26 &  22.26$\pm$  0.24 \\
       29475 &  23.43$\pm$  0.31 &  23.17$\pm$  0.25 &  22.97$\pm$  0.29 &  22.77$\pm$  0.40 \\
       29922 &   --              &  23.19$\pm$  0.23 &  22.99$\pm$  0.33 &   --              \\
       30375 &   --              &  23.11$\pm$  0.21 &  23.03$\pm$  0.29 &  22.98$\pm$  0.40 \\
       30831 &   --              &  23.01$\pm$  0.20 &  22.91$\pm$  0.30 &  22.96$\pm$  0.46 \\
       31275 &  23.70$\pm$  0.32 &  23.19$\pm$  0.23 &  23.23$\pm$  0.22 &   --              \\
       31725 &  23.77$\pm$  0.34 &  23.25$\pm$  0.23 &  23.08$\pm$  0.20 &  22.56$\pm$  0.23 \\
       32170 &  23.91$\pm$  0.31 &  23.48$\pm$  0.29 &  23.58$\pm$  0.41 &   --              \\
       32628 &  23.75$\pm$  0.33 &  23.38$\pm$  0.25 &  23.01$\pm$  0.28 &  23.18$\pm$  0.46 \\
       33077 &  23.50$\pm$  0.21 &  23.67$\pm$  0.30 &  23.26$\pm$  0.29 &  23.01$\pm$  0.29 \\
       33524 &   --              &  23.52$\pm$  0.27 &  22.99$\pm$  0.15 &   --              \\
       34369 &  24.46$\pm$  0.57 &  23.51$\pm$  0.26 &  23.83$\pm$  0.52 &   --              \\
       34815 &  24.19$\pm$  0.31 &  24.07$\pm$  0.46 &   --              &   --              \\
       35270 &  23.90$\pm$  0.31 &  23.74$\pm$  0.32 &  23.23$\pm$  0.29 &  23.44$\pm$  0.43 \\
       35715 &  24.21$\pm$  0.43 &   --              &  23.90$\pm$  0.57 &  23.11$\pm$  0.41 \\ 
\bottomrule
\end{tabular}
\label{tab:log}
\end{center}
\end{table}

\end{appendix}

\end{document}